\documentclass[usegraphicx,usenatbib]{mn2e}

\newcommand{\km}{\,\mbox{km}\,\mbox{s}^{-1}}
\def\Ha{H$\alpha$}

\title[Ionized gas outflow in NGC~4460]{Ionized gas outflow in the isolated S0 galaxy NGC~4460\thanks{Based on observations collected with the 6m  telescope of the Special Astrophysical Observatory  of the
    Russian Academy of Sciences which is operated under the
    financial support of Science Department of Russia (registration
    number 01-43)}}

\author[Moiseev et al.] {Alexei Moiseev\thanks{moisav@gmail.com},  Igor Karachentsev and Serafim Kaisin\\
Special Astrophysical Observatory, Russian Academy of Sciences, Nizhnii Arkhyz,  Karachaevo-Cherkesskaya Republic, 369167 Russia
}

\begin{document}

\date{Accepted December 29; Received 2009 December 8; in original form 2009 October 1}

\pagerange{\pageref{firstpage}--\pageref{lastpage}} \pubyear{2002}

\maketitle

\label{firstpage}

\begin{abstract}
We use  integral-field and long-slit spectroscopy to study the bright extended nebulosity discovered in the
isolated lenticular galaxy NGC~4460 during a recent  H$\alpha$ survey of nearby galaxies. An analysis
of archival SDSS, \textit{GALEX}, and \textit{HST} images indicates that current star formation is entirely concentrated in the central kiloparsec  of the galaxy disc. The observed ionized gas parameters (morphology, kinematics and  ionization state) can be explained by a gas outflow above the plane of the galaxy caused by a star formation in the circumnuclear region.
Galactic wind parameters in NGC~4460: outflow velocity, total kinetic energy -- are several times smaller comparing with the known galactic wind   in  NGC\,253, which is explained substantially lower total star formation rate.
We discuss the cause of the  star formation processes in  NGC~4460 and in two other known isolated  S0 and E galaxies of the Local
volume: NGC~404 and NGC~855. We provide evidence suggesting that feeding of isolated galaxies by intergalactic gas on a cosmological time scale is a steady process without significant variations.
\end{abstract}


\section{Introduction}

Isolated  elliptical (E) and lenticular (S0) galaxies are rather rare objects, whose  properties differ appreciably from
E and S0 galaxies in clusters and groups. Among the 513 most isolated galaxies of the Local Universe with
radial velocities $V_{LG}<3500$ only 19 systems (i.e., about 4 per cent) belong to types E and S0 \citep{Karachentsev2009}.
Isolated  E and S0 galaxies are characterized by low luminosity ($M_B\simeq -17\fm6$), bluer-than-average colors, and
appreciable presence of gas and dust. The nearby lenticular galaxy NGC~404 with $M_B=-16\fm6$,
which was found to host an extended HI shell~\citep{DelRio2004} and ultraviolet-bright   ring of young stars
\citep{Thilker2009}, is a typical example. Among  $\sim450$ Local-volume (LV) galaxies with
heliocentric distances D$<$ 10 Mpc \citep{Karachentsev2004} there are only \textbf{three} E and S0 galaxies
with absolute magnitudes [$-16.0 > M_B >-18.0$] and negative `tidal indices' TI$<$0, which allowed these
galaxies --- NGC~404, NGC~855, and  NGC~4460  --- to be classified  as isolated objects. These objects seem to be local analogues of recently identified by \citet*{Kannappan2009} `blue-sequence E/S0' galaxies which might form a transition population between  late-type spirals/iregulars and early-type red-sequence galaxies.

\begin{figure*}
\centering
\includegraphics[height=9.5cm]{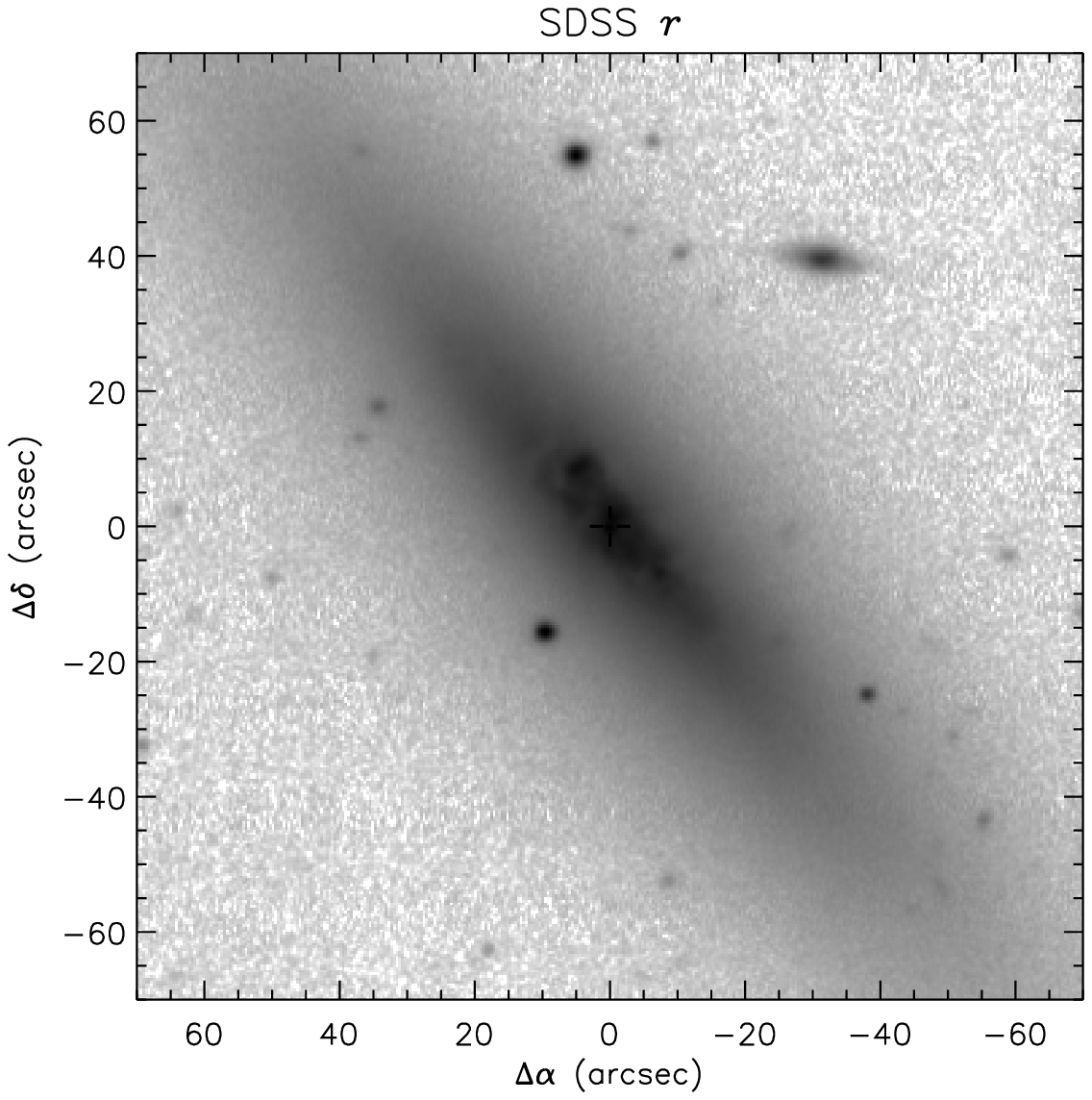}
\includegraphics[height=9.5cm]{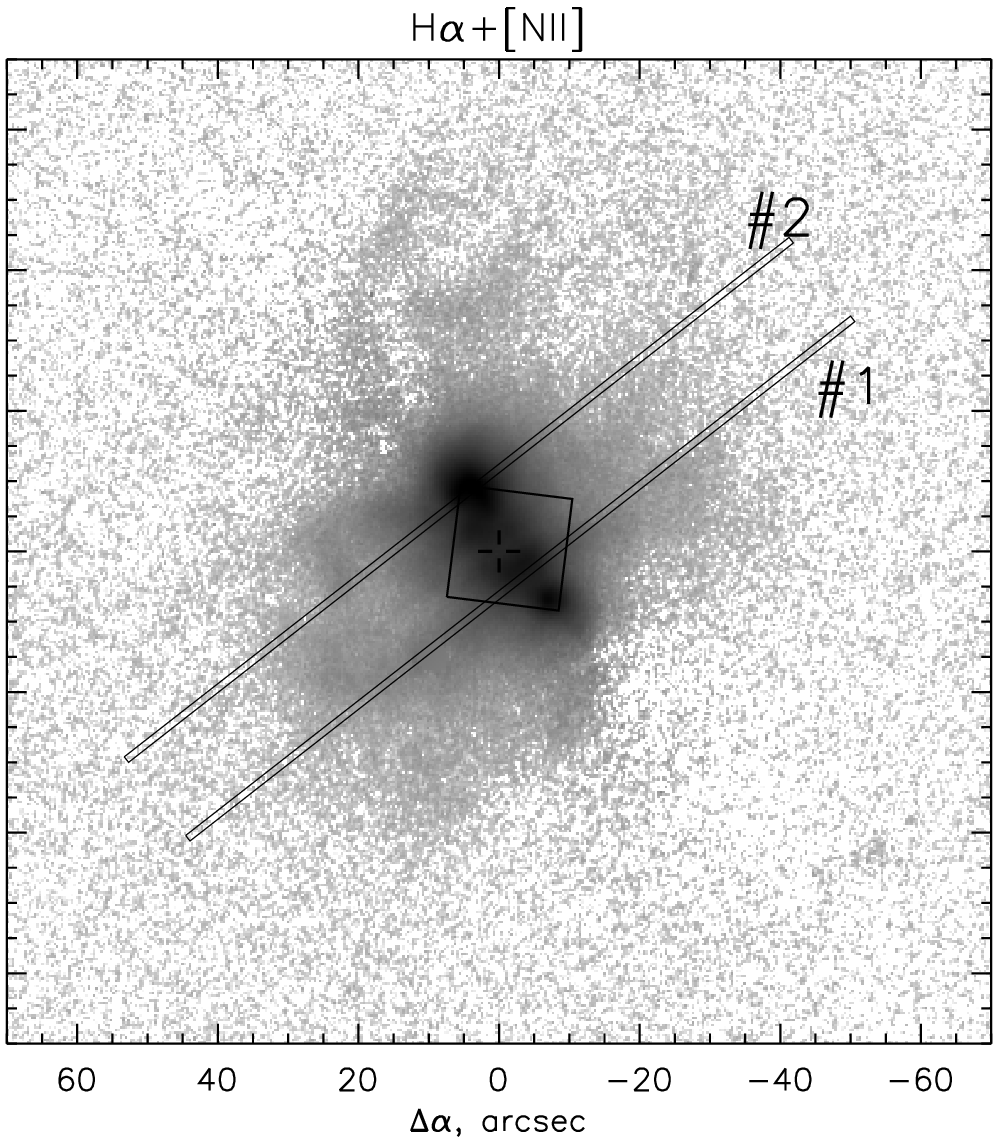}\\
\includegraphics[height=6.5cm]{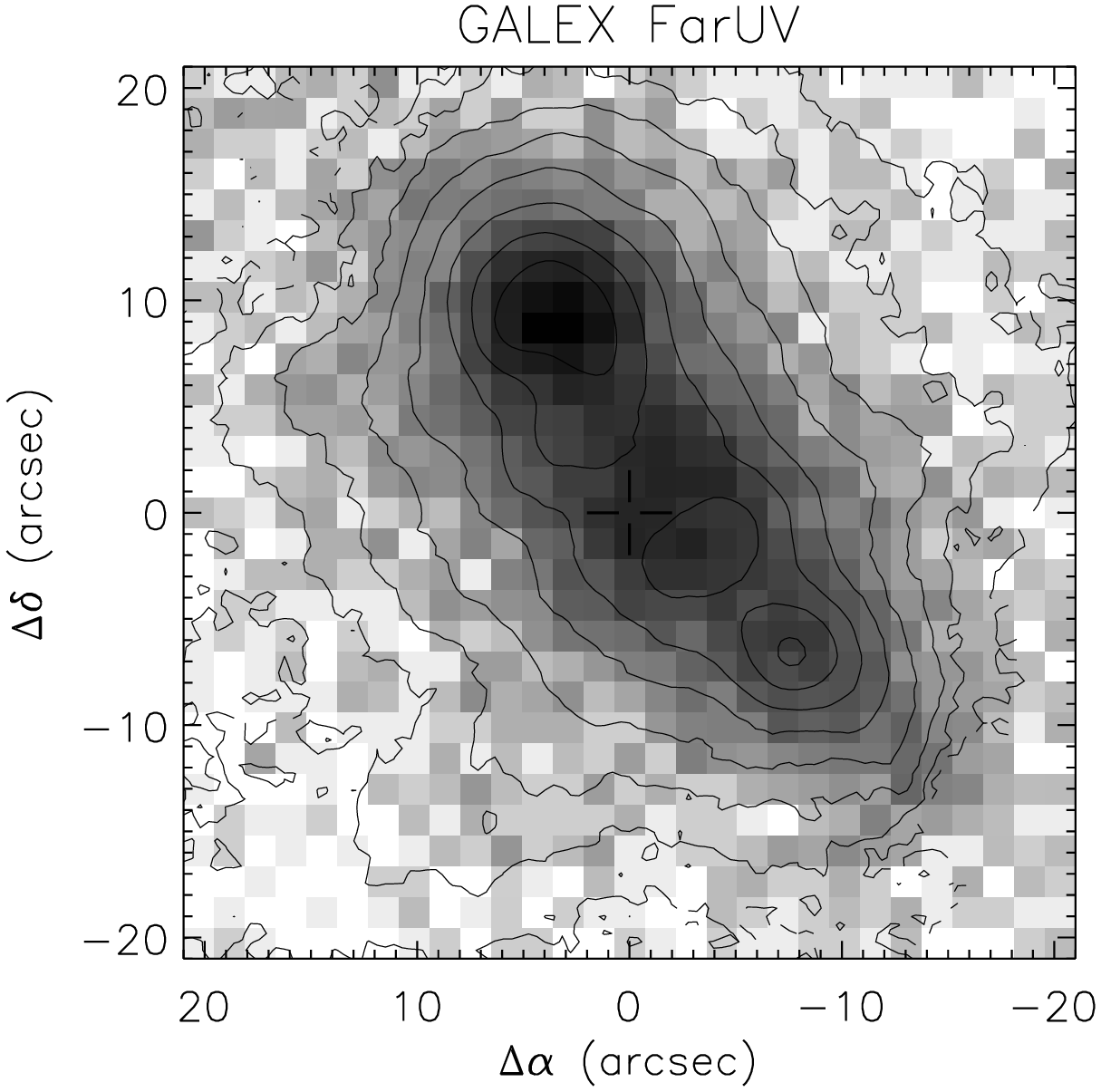}
\includegraphics[height=6.5cm]{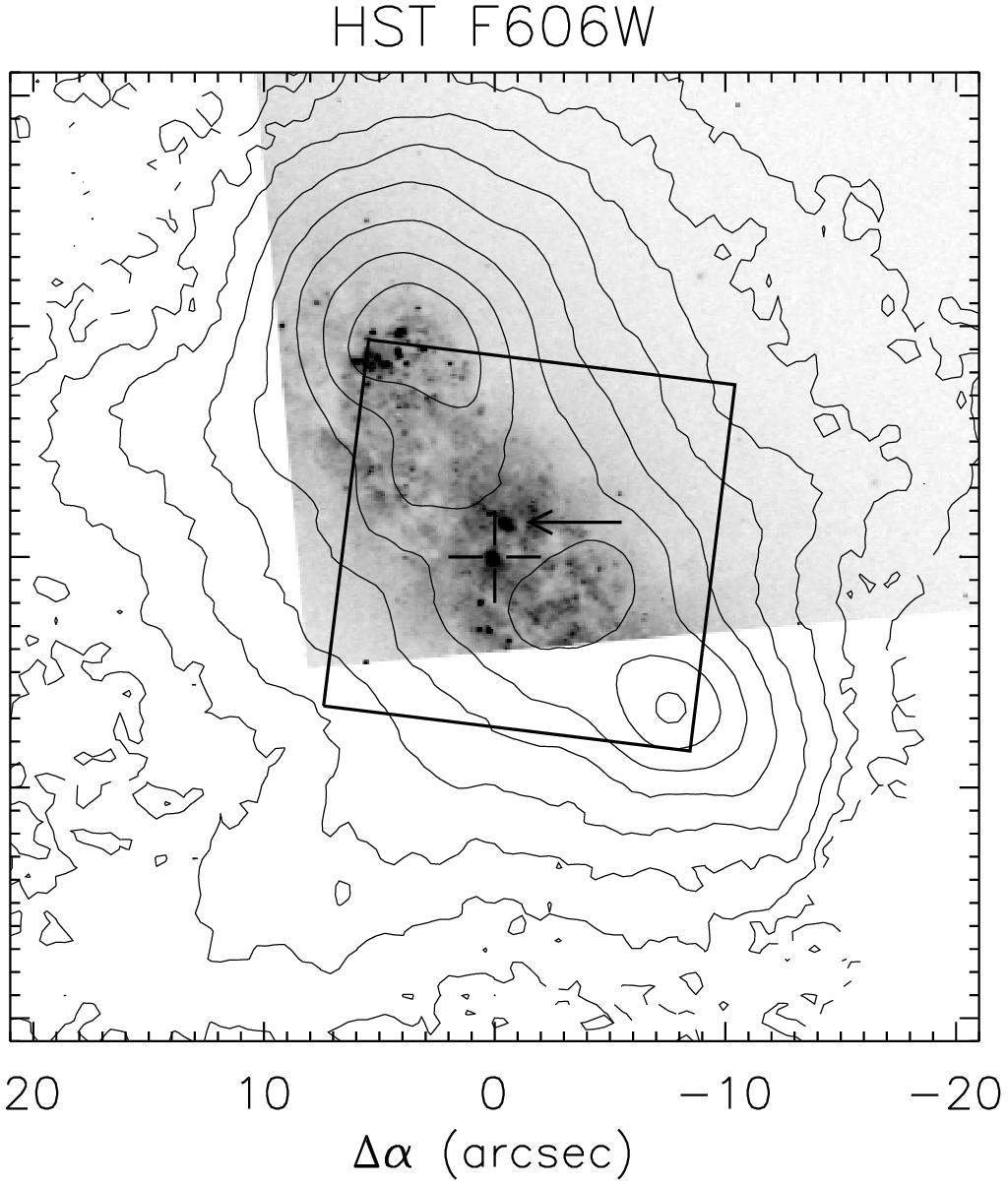}
\includegraphics[height=6.5cm]{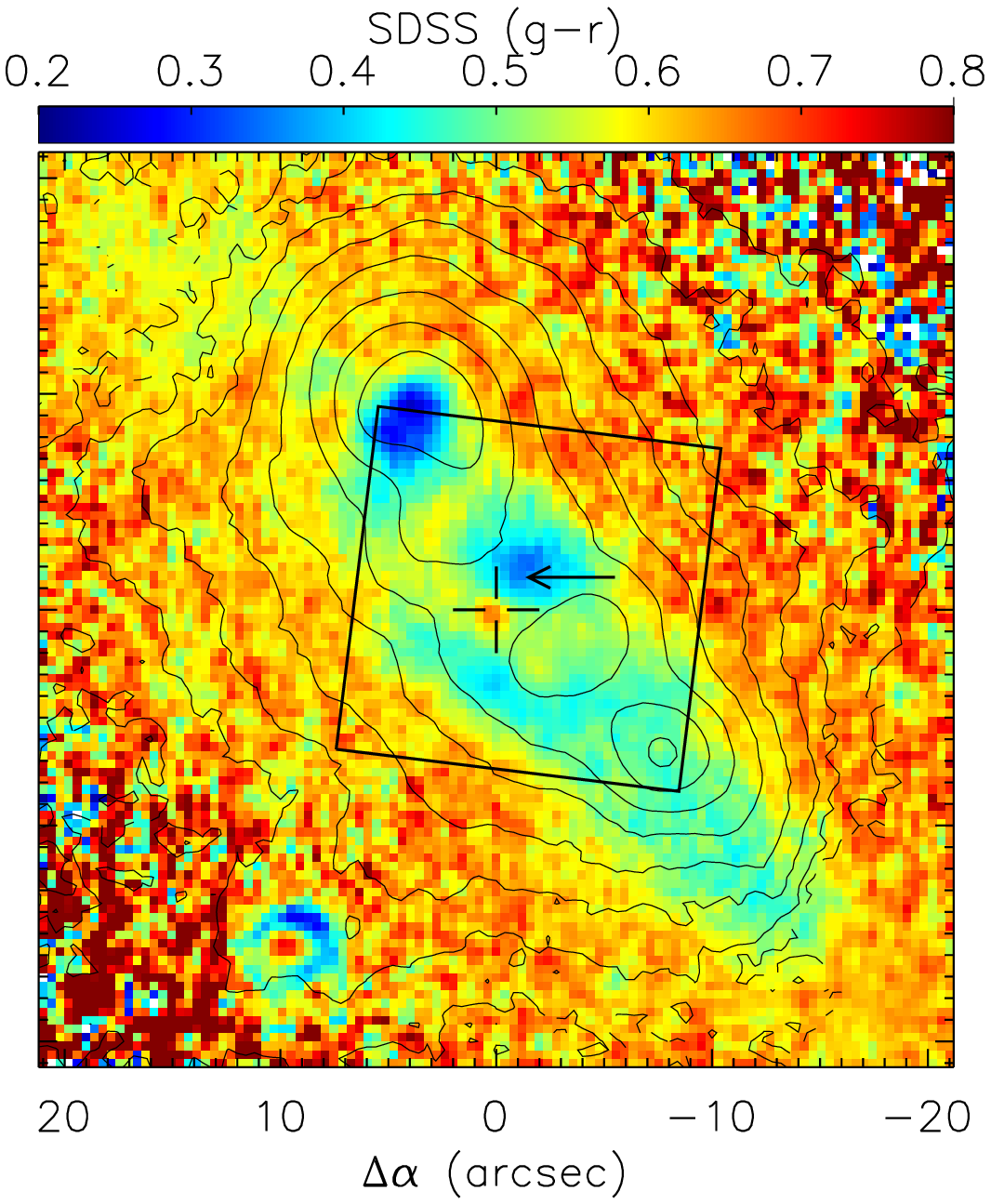}
\caption{Images of  NGC~4460 in logarithmic brightness scale. The upper row shows the SDSS $r$-band image (left) and
the \Ha+\mbox{[N\,{\sc ii}]} image (right) taken with the 6-m telescope~\citep{KaisinKarachentsev2008}.  The locations
of the spectrograph slits are shown. Hereafter the black square marks the MPFS field of view. The lower row contains the images of the center of the galaxy with \Ha\, line contours superimposed based on  \textit{GALEX}   FUV data (left),   \textit{HST} WFPC2/F606W optical map  (center)  and map of the $(g-r)$ color index according to SDSS (right). The cross marks the galaxy nucleus that is coincided with the kinematic center derived from MPFS data. The arrow  shows the position of  second circumnuclear knot (see text).}
\label{fig1}
\end{figure*}

An extensive H$\alpha$ survey of LV galaxies has been carried out in recent years with the 6-m telescope of the Special Astrophysical Observatory of the Russian Academy of Sciences (SAO RAS) in order to determine rates of star formation in a representative  distance-limited sample of galaxies. An analysis of the population of Canes Venatici~I scattered cloud  of nearby galaxies revealed  \citep{KaisinKarachentsev2008} that the circumnuclear region in  NGC~4460 hosts the bright \Ha\, line emission (see Fig.~\ref{fig1}), which must be indicative of ongoing star formation.
Such a phenomenon appears to be non-typical in an isolated early-type galaxy. This bright emission may be a result
of a recent merger of  NGC~4460 with a gas-rich companion, or of an impact of a massive intergalactic HI cloud
onto the central region of  NGC~4460. In this case, the  \Ha\, nebulosity, which extends along the major axis of the
galaxy, may be a product of a recent interaction. An analysis of the kinematics and ionization state of the gas
in NGC~4460 helps us understand the nature of this mysterious phenomenon. With this aim in view, we analyzed the available archival
images of the galaxy and performed spectroscopic observations of NGC~4460 with the SAO RAS 6-m telescope. Following  \citet{Tonry2001}, we adopt the distance to the galaxy to be 9.59~Mpc that  corresponds to a scale of  $46.5\,\mbox{pc}\,\mbox{arcsec}^{-1}$.

\section{Photometric structure of the galaxy}
\subsection{Circumnuclear region}
\label{phot1}
A comparison of the \Ha+\mbox{[N\,{\sc ii}]} image of the galaxy with broadband SDSS images (Fig.~\ref{fig1}) convincingly
demonstrates that most of the emission-line radiation emerges from a compact region in the disc inside  the
radius $r<20$ arcsec. That is where all HII regions are located, whereas diffuse  \Ha+\mbox{[N\,{\sc ii}]} emission extends along
the minor axis of the galaxy on both sides of its nucleus. An analysis of optical and UV images shows that
ongoing star formation in NGC~4460 is   concentrated entirely inside the central kiloparsec. First, it follows from the distribution of FAR-UV radiation ($\lambda1344-1786$\AA) according to \textit{GALEX} archival data,  that is exactly coincident with the circumnuclear  HII regions (Fig.~\ref{fig1}). Second, the young stellar population found in this region stands out by its bluer color index in SDSS images (see Fig.~\ref{fig1}). Star-forming regions make an almost closed ring in the disc of the galaxy. The $r$-band image reveals  a compact nucleus with $(g-r)=0.65$, whereas a condensation located $1.5-2$ arcsec North of the nucleus (marked with arrow on  Fig.~\ref{fig1})  becomes brighter in blue filters with the color $(g-r)=0.45$. On the  \textit{HST} optical image this  circumnuclear knot corresponds to  compact stellar  cluster with a linear size $FWHM=17\times10$~pc  whose luminosity is only twice lower than that of the compact cluster in the `main' nucleus.
We note that  the dynamical center of the gas velocity field (marked with cross on  Figs.~\ref{fig1} and \ref{fig3}) coincides with the main nucleus of the galaxy (see below Sect.~\ref{sec_kin}). The circumnuclear star-forming regions have evidently different evolutionary status, because only NE of the nucleus does the peak of  \Ha\, emission coincide with the  `blue' region on the  $(g-r)$ map. At the same time, the two blue regions near the minor axis  (including the second nucleus) are unassociated with \Ha+\mbox{[N\,{\sc ii}]} peaks. These regions must have already lost most of their gas or lack OB stars needed for its ionization.

The SDSS images exhibit  individual dust lanes in the inner part of the region. On the \textit{HST} image these lanes can be easily
seen to border HII regions, and they are most probably be associated with dense gas compressed by shocks produced by
star-forming processes  (SN explosions and winds from massive stars).

\begin{table*}
\caption[]{Log of spectral observations}
\label{tab1}
\centering
\begin{tabular}{lllllll}
\hline\hline
    Date       &   Device & position                            & $T_{exp}$ & sp. range  & sp. resol.   & seeing    \\
               &           &                                   &  (s)      &       (\AA)&  (\AA)       & $(arcsec)$ \\
\hline
27/28 Mar 2007 &   MPFS       & nucleus                       &   1800     &  5800-7300   &      3.5       &    1.7 \\
15/16 May 2007 &   SCORPIO/LS &PA$=128$\degr, offset 5 arcsec to SW&   1200     &  6050-7100   &      2.5       &    1.7 \\
15/16 May 2007 &   SCORPIO/LS &PA$=128$\degr, offset 9 arcsec to NE&    900     &  6050-7100   &      2.5       &    1.8 \\
\hline
\end{tabular}
\end{table*}

\subsection{Orientation of the disc and the brightness distribution}
\label{phot2}
\begin{figure}
\includegraphics[width=8.4cm]{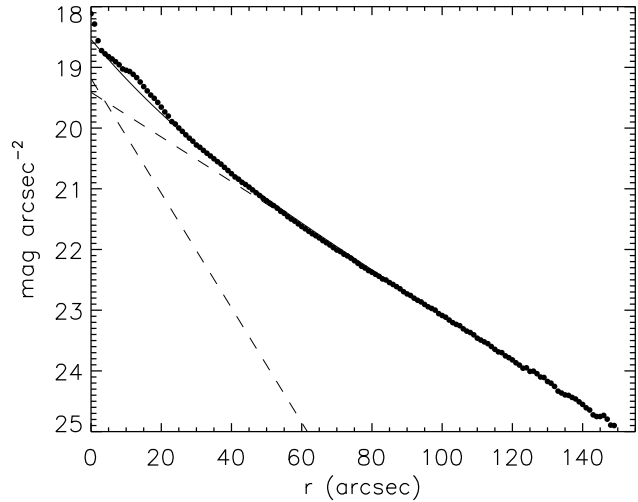}
\caption{Decomposition of the r-band surface brightness profile (dots) into two discs (the dashed line). The
solid line marks the sum of both disc models.}
\label{fig2}
\end{figure}

Using isophotal analysis of SDSS images we found the position angle and ellipticity of the outer $r$ and  $i$-band isophotes to be  $PA=39.5\pm0.4$\degr, $\epsilon=0.71\pm0.02$. According to   empirical relation \citep{Hubble1926},
this implies a galaxy inclination of $i=77\pm1$\degr. We used the inferred orientation parameters to construct the
azimuthally averaged surface-brightness profiles of  \textit{ubvri}  SDSS images.  In all filters the
brightness distribution is almost exactly exponential at $r>30-40$ arcsec, with the brightness excess at $r<20$ arcsec being more conspicuous in blue filters -- it is represented by star-forming regions.  Figure~\ref{fig2} shows decompositions
of the \textit{r}-band  surface brightness profile into components. After the subtraction of the outer disc
($\mu_0=19.4^m$, $h=29.5$ arcsec), brightness excess still remains at  $r<30-40$ arcsec. However, it is not a bulge, because after
the subtraction of the two-dimensional disc model the brightness contours of the residual image have the same
ellipticity as those of the outer disc. The brightness profile at the center can be described fairly
well by a second exponential function with almost the same central brightness ($\mu_0=19.2^m$), but with a
three times shorter scale length $h=11.5$ arcsec. At the center one can see a low-contrast compact nucleus described above.

It thus follows that only some of the morphological features  (smooth outer isophotes, lack of spiral arms)
suggest that NGC~4460 may be a lenticular galaxy, whereas the bulge/disc ratio is indicative of a later
morphological type, because the spherical component is barely visible in the brightness distribution. Inner
exponential discs are often referred to as  `pseudo-bulges' \citep{KormendyKennicutt2004}. However, the
corresponding feature in NGC\,4460 it is indeed a bona fide disc whose thickness, according to photometric data,
coincides with that of the outer disc.

Multitiered \citep*[antitruncation according the classification by][]{Erwin2005}  discs have been found increasingly often in spiral and lenticular galaxies, and they are attributed either to slow internal secular evolution that can only develop in the presence of a bar and/or a spiral~\citep[see][for review]{KormendyKennicutt2004}, or to a recent  minor merging event. Both cases imply  the loss of angular momentum by a part of gas  moved toward the center, where a burst of star formation currently occurs, and the newly born stars form the inner disc with a short scale length. Recent  formation of the inner disc is also evidenced by the fact that its contribution to the combined brightness profile (Fig.~\ref{fig2}) is almost everywhere smaller   than that of the main disc of the galaxy.

\subsection{Bar or ring?}
In Sect~\ref{phot1} we mentioned that the   star-forming regions make an almost closed ring in the central kpc. At first sight, the observed distribution of the HII regions has another interpretation, where the two brightest condensations, located almost equidistant from the nucleus $\sim10$ arcsec to north-east and south-west, belong to the ends of a stellar bar. In this case, the two blue regions appeared on the color-index map (Fig.~\ref{fig1}) at $r=2-4''$ (included the off-nuclear compact cluster mentioned above) can be considered as part of nuclear compact ring  formed on the resonances of this  bar.

However, this assumption contradicts to other morphological features. First is a significant asymmetry in the properties of two brightest HII regions. They have different size, \Ha-luminosity and color index, i.e. different history of their formation. That is strange in the case of lobes on the ends of a bar which should be formed simultaneously. Secondly, the dust in the circumnuclear region has a chaotic distribution (Sect~\ref{phot1}) instead  sharp curved dust lanes expected in a  bar.  Finally,  the  isophotal analysis of SDSS red images does not support any significant changes of ellipticity and $PA$ of isophotes at the distances corresponded to the possible bar.

The effect of non-circular  gas motions (radial inflow) could be also insufficient in the observed gas velocity field, because the  major axis of possible bar coincides with the galaxy line of nodes. Therefore, the regular velocity pattern in the circumnuclear region of ionized gas velocity field (Sect.~\ref{sec_kin}) can not be use as pro or contra arguments about bar. We think that  only data on stellar kinematics can evident  the existence (or absent) a stellar bar in NGC~4460.  However, our  analysis of galactic photometry  presented above suggests  the star-forming pseudo-ring 2 kpc in diameter nested in the inner stellar disc as more reliable  description of the observed morphology.

\section{Observations and data reduction}

Long-slit and integral-field (3D) spectral observations were made at the prime focus of the SAO RAS) 6-m  telescope. The $2048\times2048$ EEV~42-40 CCD was used as a detector in both cases.

The central region of NGC~4460 was observed with the MultiPupil Fiber Spectrograph (MPFS). The MPFS \citep*{Afanasiev2001}  takes simultaneous spectra from 256 spatial elements (constructed in the shape of square lenses) that form on the sky an array of $16\times16$ elements (`spaxels') with the angular size $1\,\mbox{arcsec}\,\mbox{spaxel}^{-1}$.  A bundle of 17 fibers placed at the distance about 3.5 arcmin  from the lens array provides the night-sky background spectra simultaneously with objects exposition. The spectral range included  bright emission lines of ionized gas: \Ha, \mbox{[N\,{\sc ii}]}$\lambda\lambda6548,6583$  and  \mbox{[S\,{\sc ii}]}$\lambda\lambda6717,6731$.  Fig~\ref{fig1} shows the locations  of MPFS lens array on the galaxy image, the log of observations is presented in Table~\ref{tab1}.
The spectra were reduced using the IDL-based software,  the data reduction sequence is briefly described in \citet*{Moiseev2004}.  The data reduction results in a data cube in which a spectrum corresponds
to each image `spaxel'  in two-dimensional field  $16\times16$  arcsec$^2$. The data cube was flux-calibrated using the standard star observations taken on the same night, just after the galaxy. The maps and velocity fields of the H$\alpha$, \mbox{[N\,{\sc ii}]} and \mbox{[S\,{\sc ii}]} emission lines were constructed using a single-Gaussian fitting.

The long-slit observations have been made with the multi-mode focal reducer SCORPIO \citep{AfanasievMoiseev2005}. We obtained two cross-sections in direction near perpendicular to the galactic major axis, along extended  \Ha\,filaments and though bright HII-regions in the galaxy disc (hereafter slits \#1 and \#2, see Fig.~\ref{fig1}). The slit width was $1$ arcsec, the scale  along slit was $0.35\,\mbox{arcsec}\,\mbox{px}^{-1}$,  the log of observations is given in Table~\ref{tab1}. The slit spectra were reduced and calibrated using the IDL-based software, developed in SAO RAS. The parameters of the emission lines (FWHM, flux, velocity) were calculated from a single-Gaussian fitting. A multicomponent structure of the emission line profiles was not detected as well as in MPFS spectra. The long-slit spectra are deeper, because  SCORPIO quantum efficiency  is significantly higher than MPFS. This fact allows us to detect the weak emission lines [OI]$\lambda\lambda6300,6360$ and He~I$\lambda6678$ in the brighter HII regions in the galactic disc.

\section{Parameters of ionized gas}

\begin{figure}
\centering
\includegraphics[height=4.5cm]{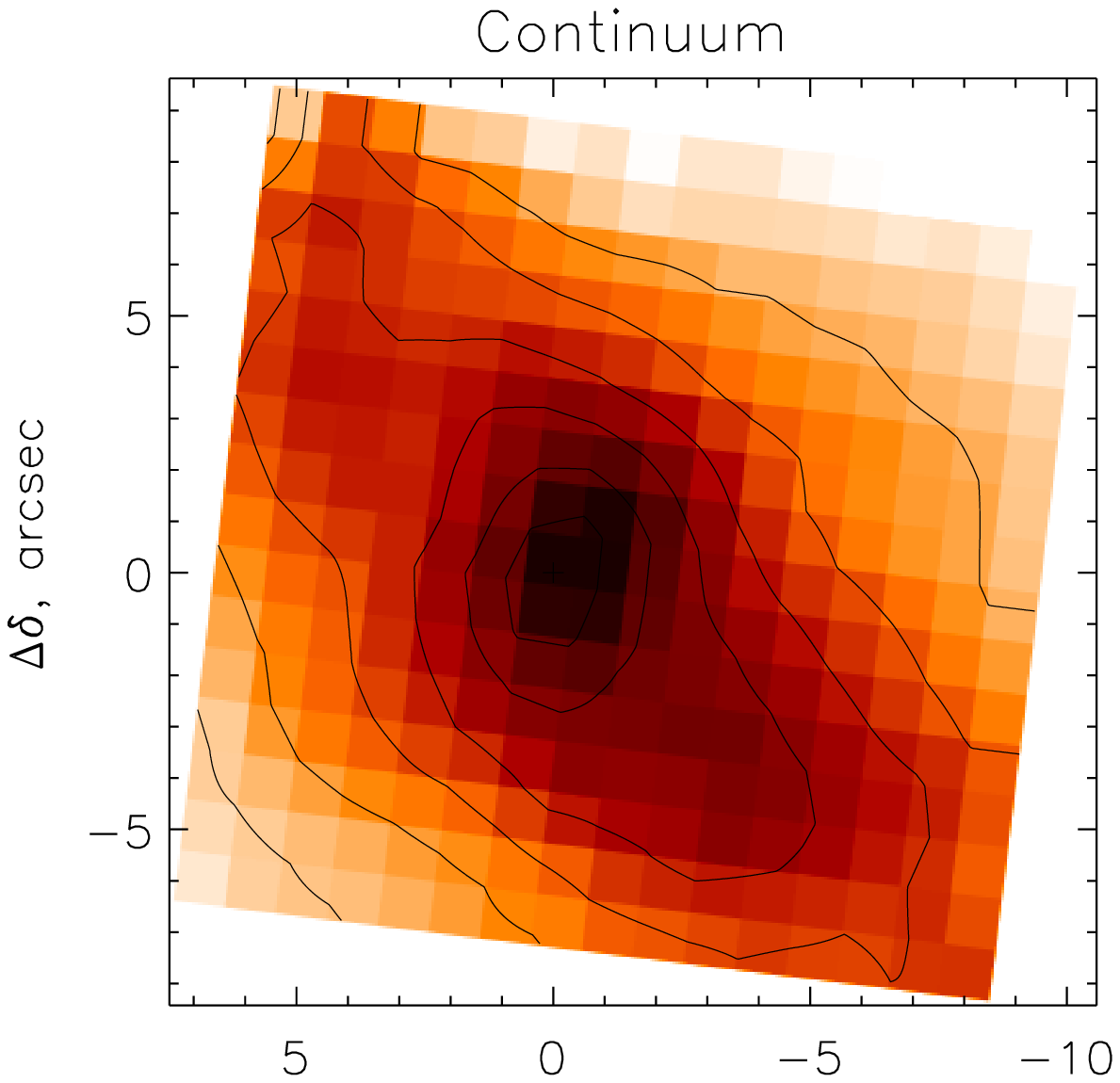}
\includegraphics[height=4.5cm]{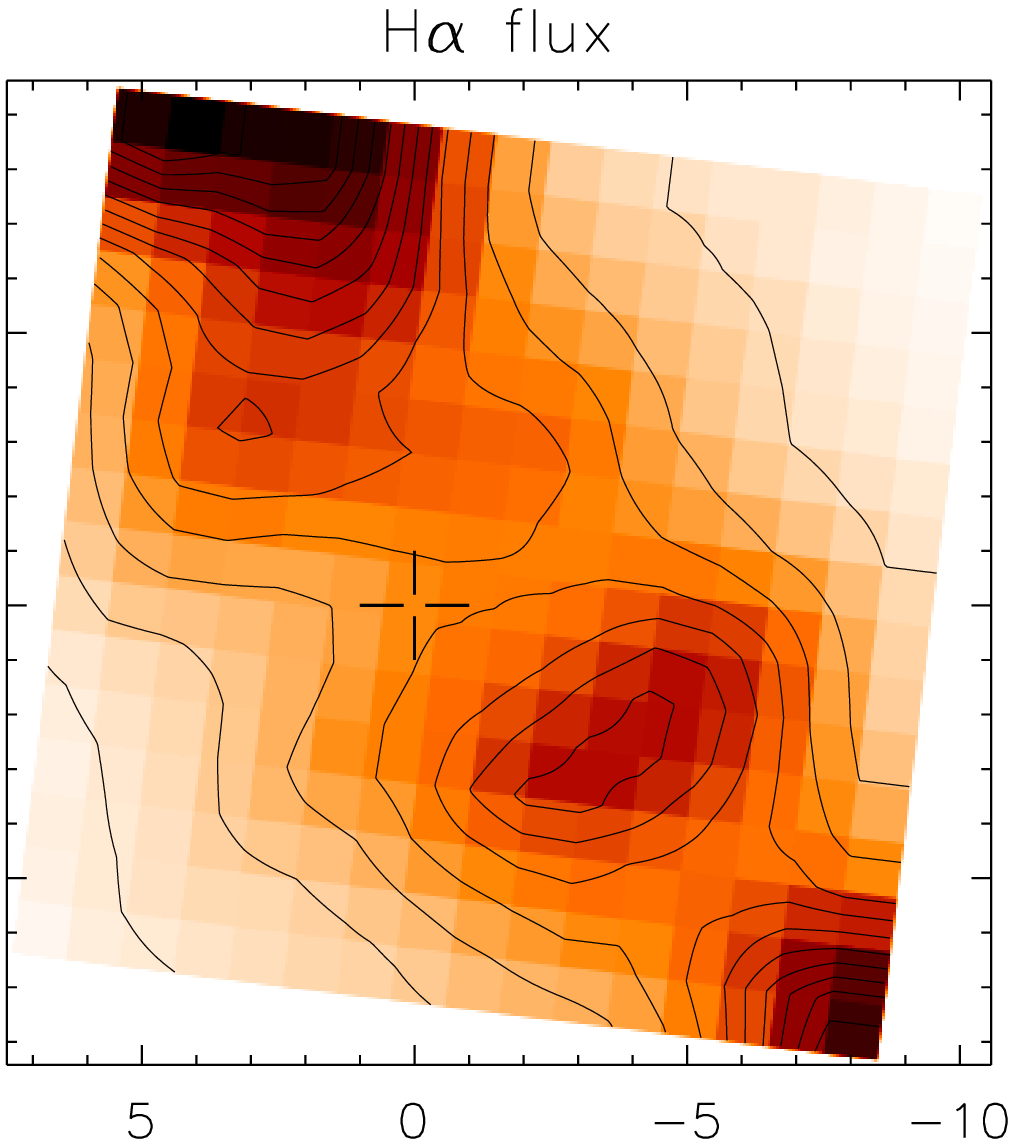} \\
\includegraphics[height=4.5cm]{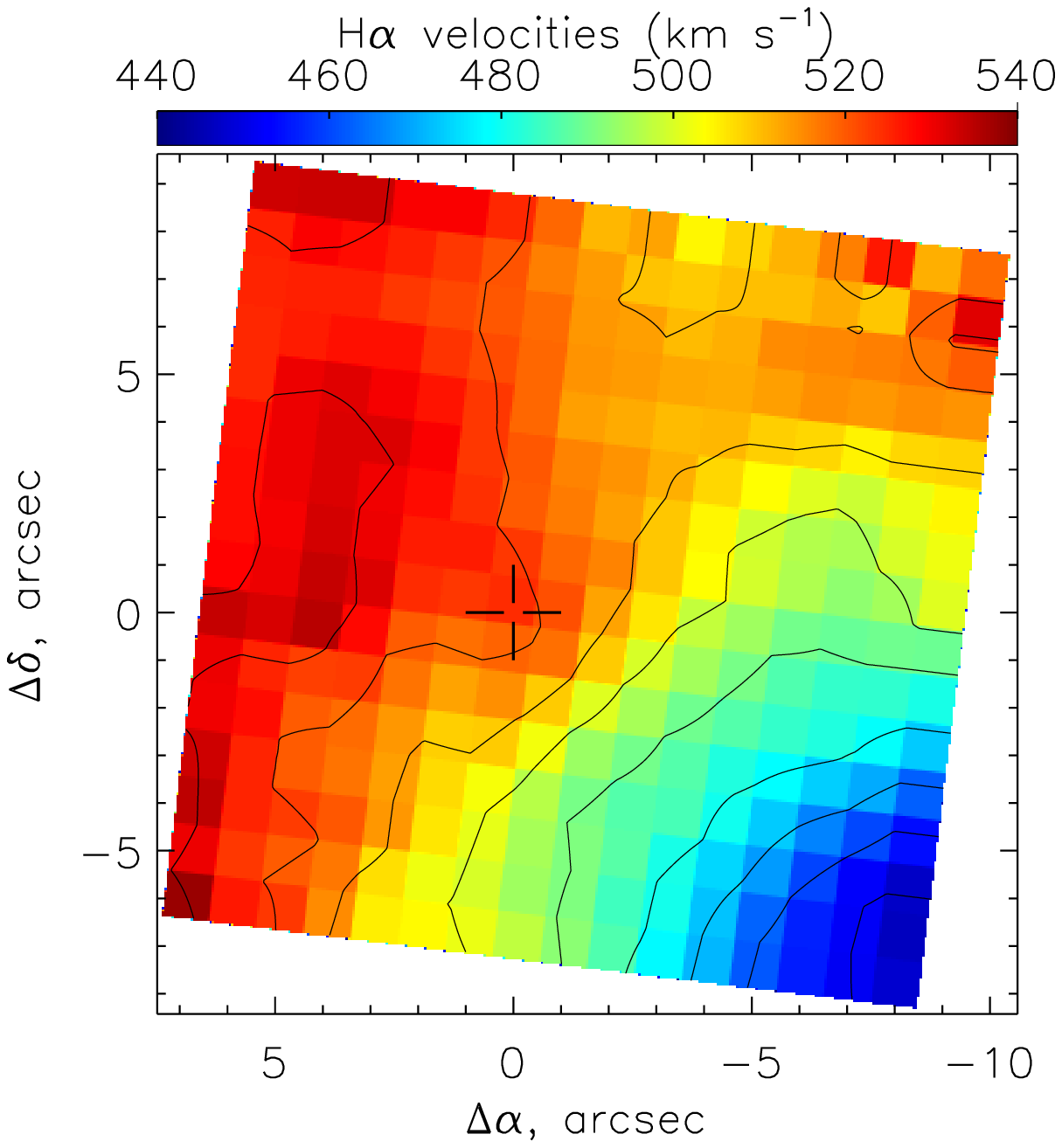}
\includegraphics[height=4.5cm]{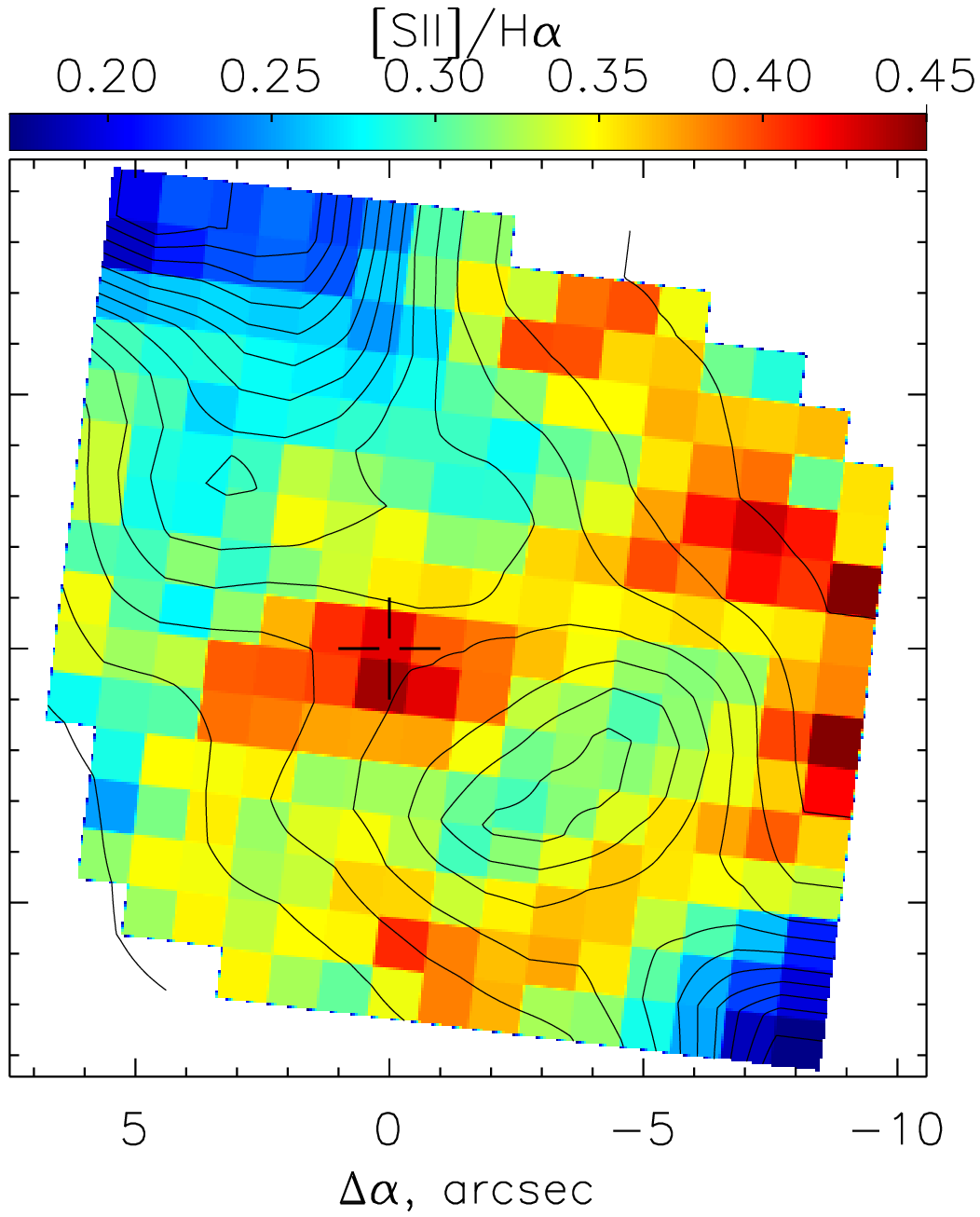}
\caption{Results of MPFS observations. The top panel shows the continuum image taken near $\lambda6400$\AA (left)
and the \Ha\, map (right). The bottom panel shows the  \Ha\, velocity field (left) and the map of the
\mbox{[S\,{\sc ii}]}$(\lambda6717+\lambda6731)$/H$\alpha$ flux ratio with  \Ha\, contours superimposed (right). The coordinate
origin coincides with the dynamic center (the cross).}
\label{fig3}
\end{figure}

Fig.~\ref{fig3} shows several maps based on MPFS data. The \Ha\, brightness distribution agrees well with the
narrow-band filter image discussed above. The continuum image exhibits a central condensation extending along the N-S
direction. Given the lower spatial resolution of the MPFS compared to SDSS data, the condensation must correspond
to the double nucleus described in Sect.~\ref{phot1}. The MPFS field of view  contains three bright HII regions.  The maps of the  \mbox{[N\,{\sc ii}]}/\Ha\, and \mbox{[S\,{\sc ii}]}/\Ha\, flux ratios
indicate a slight enhancement of forbidden-line emission around star-forming regions\footnote{Hereafter
by the  \mbox{[N\,{\sc ii}]}/\Ha\,and  \mbox{[S\,{\sc ii}]}/\Ha\, ratios we mean the  \mbox{[N\,{\sc ii}]}$\lambda6583$/\Ha\ and
\mbox{[S\,{\sc ii}]}$(\lambda6717+\lambda6731)$/\Ha\, line flux ratios, respectively.}, whereas the corresponding ratios are minimal
at the centers of HI regions (see Fig.~\ref{fig3}). In the space between star-forming regions we most likely observe
hot gas ejected from these regions and ionized, among other factors, by shocks. The presence of shock fronts around HII
regions is further evidenced by dust lines seen at these locations (Sect. \ref{phot1}). In these places the
\mbox{[S\,{\sc ii}]}/\Ha\, ratio reaches  0.45, which is close to the usually adopted border between shock ionization and
photoionization (see below).

Figure~\ref{fig4} shows variations of the emission-line parameters (flux, radial velocity, $FWHM$ corrected
for instrumental broadening) along the slits. The maximum of the emission lines brightness is therefore shifted relative to the peak in brightness of stellar continuum, because bright HII regions in the circumnuclear ring do not lie on the major axis of the galaxy. Beyond the disc, surface brightness in the
nebulosity decreases almost exponentially and emission lines can be detected at even greater distance from the disc than
on the  \Ha+\mbox{[N\,{\sc ii}]} image. The \mbox{[S\,{\sc ii}]}/\Ha, intensity ratio increases with increasing distance from HII regions, whereas
the corresponding increase of the  \mbox{[N\,{\sc ii}]}/\Ha\, ratio is less appreciable (see Fig.~\ref{fig4}).

To identify the source of gas ionization, we constructed the diagnostic diagram showing the ratios of the fluxes
of lines with different excitation mechanisms and similar wavelengths so that the result would not depend on
dust extinction. Unfortunately, we can construct only one such diagram in the wavelength interval considered:
\mbox{[N\,{\sc ii}]}/\Ha\,versus \mbox{[S\,{\sc ii}]}/\Ha\, (Fig~\ref{fig5}). The dashed lines separate regions with different excitation mechanisms
indicated as `photoionization' and `shock'. We assume, in accordance with \citet{Stasinska2006},
that in the case of ionization by young stars the line rations  follow to relations:
$\log\mbox{[S\,{\sc ii}]}/\mbox{H}\alpha<-0.4$, $\log\mbox{[N\,{\sc ii}]}/\mbox{H}\alpha<-0.4$. Higher relative intensity of forbidden lines means that
shocks, or even a nonthermal source (AGN), contribute appreciably to ionization. When interpreting the positions of
data points on the diagram one must bear in mind that: (1) boundaries of the photoionization domain are somewhat
controversial. Namely, the boundaries  are determined not only by particular models, but also by the statistics of observations of different galaxies samples; (2) in real objects the  combined effect of several ionization sources is present. Thus, according to
\citet{Stasinska2006}, the  $-0.2<\log\mbox{[N\,{\sc ii}]}/\mbox{H}\alpha<-0.4$ domain corresponds to a composite source of ionization in LINER-type galaxies: photoionization by stars plus the effect of shocks, and shocks
dominate only at $\log\mbox{[N\,{\sc ii}]}/\mbox{H}\alpha>-0.4$. Note that many researchers use even harder criterion
\citep*[e.g., $\log\mbox{[N\,{\sc ii}]}/\mbox{H}\alpha>0$ in][]{Veilleux2005}.

Figure~\ref{fig5} shows that the gas in the circumnuclear region (i.e. central part  observed with the MPFS, and the  $r<5$ arcsec regions in slit \#1 and $r<10$ arcsec regions in slit \#2)  is ionized by
young stars. In the outer filaments shock ionization begins to dominate with increasing distance from the disc plane, according to the \mbox{[S\,{\sc ii}]}/\Ha\, criterion. At the same time,  the $\mbox{[N\,{\sc ii}]}/$\Ha\, criterion indicates that
all our measurements, except several data points with $r\approx30$ arcsec  in slit \#2, lie in the photoionization
domain. However, in this case we consider the latter criterion to be less important, because the observed line ratio
agrees well with the results of model computations~\citet{Allen2008} for shock+precursor ionization with number density
$n=1\,\mbox{cm}^{-3}$ and solar metallicity. Figure~\ref{fig5} shows the grid of parameters for this model. It is
evident from this figure that with increasing  $r$ the observed data points lie along the lines of the increasing
shock velocity from $200$ to $300\km$ for small values of magnetic parameter $0-1\,\mu G\,\mbox{cm}^{2/3}$.  For  estimation of gas mettalicity  we used the data of slit \#2 that passes through the center of the brightest HII  region, where the $\mbox{[N\,{\sc ii}]}/\mbox{H}\alpha$ ratio is minimal and equal to $0.16\pm0.03$. If we assume that the shock contribution is minimal at this location then the approximation of \citet{Stasinska2006} implies a gas metallicity of $0.74\pm0.04\,Z_\odot$, i.e. the solar-metallicity approximation is quite applicable to  NGC~4460.

It thus follows that gas ionization in the extended  \Ha\, nebulosity can be explained by the combined effect of shocks and
photoionization by young stars with the latter dominating in the inner regions. The galaxy lacks an active nucleus and
therefore the likely cause of shocks is star formation.  Gas is blown above the plane of the galaxy by
the combined action of supernovae  and stellar winds from young massive stars. This phenomenon is known as a
`galactic wind' \citep[see a review by][]{Veilleux2005}. For comparison, we present in Fig.~\ref{fig5}
the line ratios for three well studied galaxies where the existence of galactic wind has long been proven:
M~82, NGC~253, and NGC~1569. For M~82 we  give two  average values obtained by \citet{Westmoquette2009}, who used  DensePak and GMOS instruments. We adopt the galactic-wind data for
NGC~253 from Fig.~5 in  \citet{Matsubayashi2009}, and for NGC~1569 we give several estimates for external emission
regions adopted from Fig.~7 in \citet{Heckman1995}, for slit $PA=70^\circ$ and $r=\pm20,\pm50,\pm70$ arcsec from the center. On the diagnostic diagram the published data for these well-known galactic winds scattered in  relative wide range ($\pm0.5$~dex), which also includes the data points for  NCC~4460 . Shock contribution to gas ionization in the
filaments of NCC 4460 is comparable to galactic winds  for all three galaxies hitherto studied. Moreover, the $\mbox{[S\,{\sc ii}]}/\mbox{H}\alpha$ ratio in NGC~4460 is comparable with the values observed in NGC~253 and NGC~1569. That is not for $\mbox{[N\,{\sc ii}]}/\mbox{H}\alpha$, because the intensity of the nitrogen doublet  also depends from from metallicity.  For instance,  $\mbox{[N\,{\sc ii}]}/\mbox{H}\alpha$ ratio in NGC253 is most likely due to the nitrogen overenrichment of the interstellar medium~\citep{Matsubayashi2009}.

We use the \mbox{[S\,{\sc ii}]} line doublet ratio to estimate the electron density $n_e$ in accordance with the relation of
\citet{Osterbrock1989} for $T_e=10\,000\,\mbox{K}$. Figure~\ref{fig6} shows the radial variations of of the line ratio and corresponded values of $n_e$.
Unfortunately, Osterbrock's relation can be applied with confidence only at $n_e>50-100\,\mbox{cm}^{-3}$,
whereas at low densities the slope of the  (\mbox{[S\,{\sc ii}]}$\lambda6717/\lambda6731$ vs $n_e$) relation  flattens and
therefore leads to large uncertainties. Although the observed line ratio lies mainly inside this `inconvenient' interval, it is safe to conclude that the density of emitting gas is low than $100\,\mbox{cm}^{-3}$ in the disc of the galaxy. Outer
filaments exhibit a large scatter of density values ranging from zero to $200-300\,\mbox{cm}^{-3}$.  It might be supposed that in
the outer regions the spectrograph slit crosses individual clumps with highly nonuniform density distribution.

\begin{figure*}
\centering
\includegraphics[width=16cm]{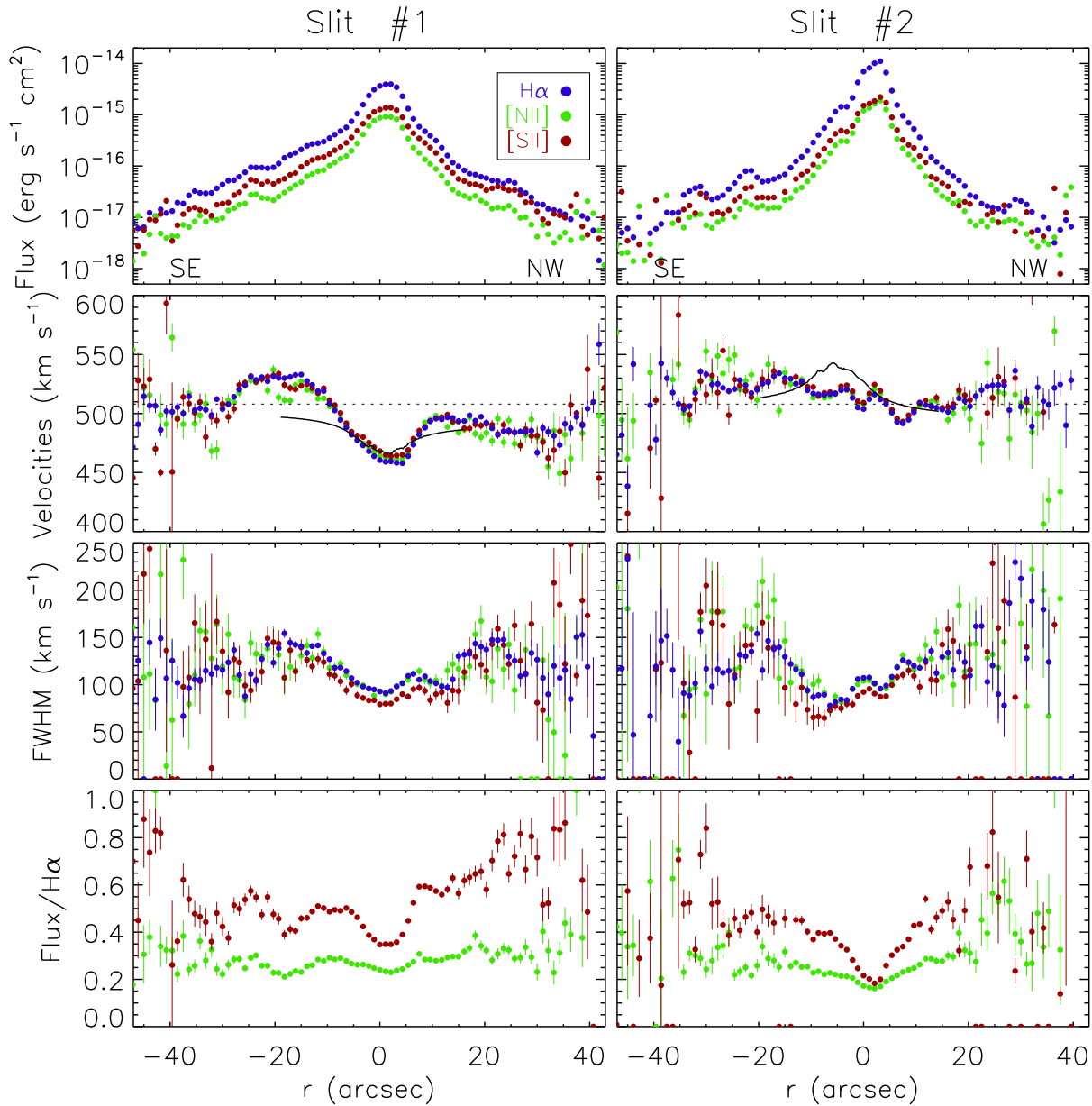}
\caption{Measurements of parameters of the \Ha, \mbox{[N\,{\sc ii}]}$\lambda6583$ and \mbox{[S\,{\sc ii}]}$\lambda6717+6731$ emission lines
along the slits \#1 (left) and  \#2 (right). From top to bottom: surface brightness, line-of-sight velocity,
FWHM (corrected for the widths of the instrumental profile), and the line-to-\Ha\, flux ratio. The dashed and solid
lines show the systemic velocity (according to MPFS data) and the cross section of the model velocity field,
respectively.}
\label{fig4}
\end{figure*}

\begin{figure*}
\centering
\includegraphics[width=17cm]{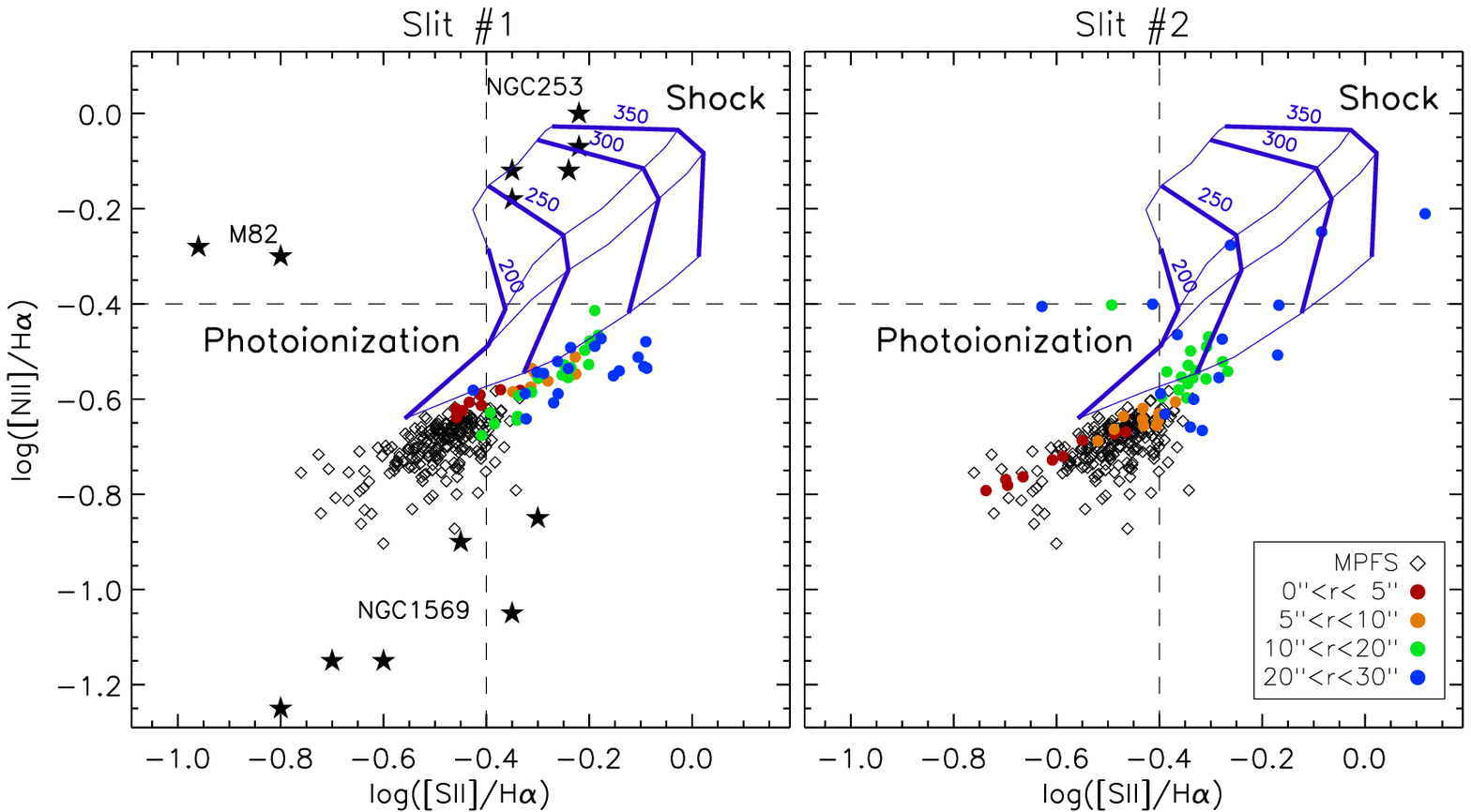}
\caption{The diagram of the \mbox{[N\,{\sc ii}]}/\Ha\,vs \mbox{[S\,{\sc ii}]}/\Ha\, flux ratios. The dashed line separates domains with different ionization mechanisms. The blue lines show the grid of shock+precursor  ionization models according to \citet{Allen2008} for $n=1\,\mbox{cm}^{-3}$  and solar elemental abundances.  The thin and bold blue lines mark the contours of constant  magnetic parameter $0.001,0.5,1,5\,\mu G\,\mbox{cm}^{2/3}$ (from bottom  to top) and contours of  constant  shock velocity (which is labeled in $\km$), respectively. The circles show the measurements made along slits \#1 (left) and \#2 (right). Different
colors correspond to different distances along the slits. The diamond signs indicate the results of MPFS measurements for the circumnuclear region. The asterisks in the
left-hand figure show the measurements for the regions of galactic wind in  M~82 \citep{Westmoquette2009}, NGC~253 \citep{Matsubayashi2009}, and NGC~1569 \citep{Heckman1995}.
}
\label{fig5}
\end{figure*}

\section{The gas kinematics}
\label{sec_kin}

The H$\alpha$-line velocity field based on MPFS measurements (Fig.~\ref{fig3}) can be described fairly well
in terms of the model of circular rotation of the thin disc. The line-of-sight  velocities measured by the \mbox{[N\,{\sc ii}]} and \mbox{[S\,{\sc ii}]}
lines show no appreciable deviations from  \Ha\, line-of-sight velocities. The center of rotation defined as
the symmetry center of the velocity field coincides with the center of the continuum image.

We constructed the model of rotation using the method described by \citet{Moiseev2004} with the position angle and
inclination  of the disc fixed in accordance with photometric data (Sect. \ref{phot2}). The $PA$ measured from the velocity field agrees with the photometric position angle within the errors. The systemic velocity  is $V_{sys}=508\pm\km$. The rotation curve shows solid-rotation increase up to $v_{max}=67 \km$
at $r=20$ arcsec that is a border of MPFS field-of view. This value is close to the maximum rotation velocity according to HI measurements~\citep{Sage2006}
 Observed velocities deviate from the corresponding model values by up to $20\km$ along the minor axis of the galaxy, and this behavior is most probably indicative
of gas ejection above the plane of the galaxy. It could also be due to the uncertainties  in the model: the thin-disc approximation may perform poorly at $i=77\degr$, internal absorption should be taken into account, etc.

The ionized gas velocity field  in the  compact star cluster region  noted on optical images at $\sim2$ arcsec North  of the nucleus (see Sect.~\ref{phot1}) does not show any kinematic peculiarities, like non-circular motions or shift of the rotation center. Therefore, our data do not support that this cluster is a nucleus of a dwarf satellite -- a merging remnant. New high-resolution stellar kinematics data are needed for  understanding the origin of this off-set cluster.

To estimate the peculiar velocities of the outer filaments, we extrapolate our model out to greater distances assuming
that the rotation curve reaches a plateau.  Figure~\ref{fig4} compares the model extrapolation and observed  velocities
in slit cross sections. Slit \#1 shows a good agreement between the model and observations at $r<15$ arcsec in the NW half of the nebulosity, whereas its more extended SE half exhibits appreciable deviations already beyond
$7-10$ arcsec from the center with the discrepancies amounting to $30-35\km$. Slit  \#2 passes farther from
the center, along the edge of the field of view of the MPFS, and therefore here deviations from the model
extrapolation are more conspicuous in the disc region. On the whole, we may conclude that in the outer filaments
systematic deviations of the observed line-of-sight  velocities from the corresponding model values do not exceed, on the
average,  $30\km$. This implies  an outflow velocity of $v_{out}\approx130\km$, if we assume that the observed peculiar gas motions are directed perpendicularly to the plane of the disc and take projection effect into account.
This value is comparable to the observed halfwidth of emission lines, which amounts to $150-170\km$ at a distance
of $r=20-30$  arcsec (0.9-1.4 kpc), whereas the FWHM is half in the disc. The fact that the amplitudes of regular
velocities are approximately equal to those of chaotic velocities are  indicative of strong turbulence in the filament gas.

\section{Discussion}
\subsection{Galactic wind}

We showed above that the peculiarities of the distribution of the parameters of ionized gas (kinematics, density, and
state of ionization) observed in NGC~4460 can be most easily explained in terms of the `galactic wind' hypothesis,
i.e., assuming gas outflow above the plane of the galaxy produced by a central burst of star formation.
The alternative assumption suggesting that we observe  remnants of a tidally disrupted gas-rich companion is ruled out,    because it would imply that line-of-sight  velocities in filaments should deviate significantly from the systemic
velocity. Besides the above indicators, the  shape of the emission nebulosity also suggests the presence of galactic
wind. The nebulosity resembles a butterfly or, rather, two bubbles expanding above the plane of the galaxy. Note that
the diameter of the bubble has become greater than the halfwidth of the gaseous disc, it has open vertex, because the latter is no longer compressed by the ambient gas. Moreover, the emission nebulosity resembles  the well-known superwind galaxy NGC~1482 \citep[see review by][for this and other examples]{Veilleux2005}.
The nonuniform  distribution of the density in the emission nebulosity  is  not surprising, because the optical emission lines are thought to originate from  a boundary between hot bubble gas and cool ambient ISM, fragmented under the action of Kelvin-Helmholtz and Rayleigh-Taylor instabilities.

Regions of modern star formation are located at the base of the wind  and we showed that ongoing star formation
is concentrated almost entirely inside a ring less than 2~kpc in diameter. Gas is ionized not only by the radiation
of OB stars, but also by shocks produced by the combined effect of supernovae and winds of
massive stars, which transfer kinetic energy to the interstellar medium. The observed line-of-sight velocities in
filaments are small and close to the systemic velocity, suggesting that these features consist of gas that was ejected
from central regions and has preserved its angular momentum.The gas outflow occurs mostly in the sky plane and therefore the line-of-sight projection of the outflow
velocity is small.

\begin{figure}
\centering
\includegraphics[width=8.4cm]{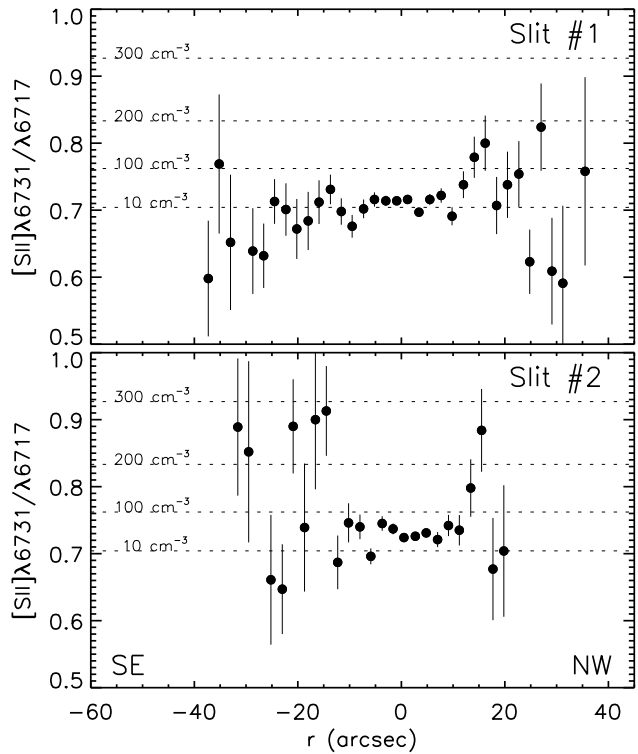}
\caption{Variation   of the \mbox{[S\,{\sc ii}]} line doublet  ratio  for slit~\#1 (top) and  slit~\#2 (bottom). The horizontal dashed lines indicate the values of electron density corresponded which are labeled near the lines.}
\label{fig6}
\end{figure}

We  adopt the characteristic length of the brightest filaments $l=30-35$ arcsec (1.3-1.6 kpc) to derive the dynamic
age of the emission feature $\tau_{dyn}=l/v_{out}\approx10-12\,$Myr. The total  \Ha\, luminosity of the galaxy
is $L_{H\alpha}=1.7\times10^{40}\,\mbox{erg}\,\mbox{s}^{-1}$ based on the observed \Ha+\mbox{[N\,{\sc ii}]} flux from
\citet{KaisinKarachentsev2008} and the average line ratio of $\log\mbox{[N\,{\sc ii}]}/\mbox{H}=-0.7$. We assume that
the wind radiation includes the flux from the regions located outside the  ellipse with the
semi-major axis  $25$ arcsec and axial ratio 0.4, which includes all HII regions in the disc. The wind
luminosity  is  $L_{H\alpha}(wind)=2.7\times10^{39}\,\mbox{erg}\,\mbox{s}^{-1}$, it makes up for 16 per cent of
the total  $L_{H\alpha}$, and the SE and NW parts of the bipolar structure have almost equal luminosities (9
and 7 per cent, respectively). We now adopt, in accordance with Fig.~\ref{fig6}, a mean electron density of
$n_e=50\,\mbox{cm}^{-3}$ to infer, in the same way as \citet{Matsubayashi2009}, the total mass of ionized
gas ejected from the disc\footnote{In this analysis we used  information about single  phase of the wind, because \Ha\, emission traces only the cooler interaction zone between hot wind fluid and the ambient medium. Therefore the  values of $M_{wind}$ and $E$ are underestimated. However, we present its for comparison with the similar values taken from similar analyses in other galaxies.}, $M_{wind}=1.7\times10^5\,M_\odot$, and its kinetic energy $E=5.8\times10^{52}$.
We describe the emitting volume of the  SE part  of the outflow wind by a truncated cone with the bases of diameters 30 and
40 arcsec to derive a characteristic filling factor for this volume  $f\approx3\times10^{-5}$. This value of $f$  is indicative
of the highly nonuniform nature of the emitting medium --  \Ha\, emitting gas is located mostly in the walls
of the bubble, which fragment under the action of characteristic instabilities \citep{Veilleux2005}.

We now compare our inferred parameters with the wind parameters in two well-known nearby galaxies: M82
\citep{Shopbell1998} and NGC~253 \citep{Matsubayashi2009}. The mass of the ejected gas in NGC~253 is almost the same
as in our case, although the kinetic energy of the gas is twice  lower, whereas the formation time scale of the
feature is almost six times shorter. In M82 $M_{wind}$ and $E$ are greater by almost a factor of  30 and 360,
respectively, with closer $\tau_{dyn}=3.4$ Myr. The wind outflow velocities in these galaxies are three to four times
higher than in NGC~4460. Given our remark about the uncertainty of measured $v_{out}$, it can be concluded that,
on the whole, the wind parameters in NGC~4460 are close to those observed in NGC~253, except for the fact that
the emission nebulosity in  NGC~4460 took much longer time to form. This is not  surprising, because
the total  SFR formally inferred from $L_{H\alpha}$ is almost 10 times higher in NGC~253. This fact  most likely
determines the relatively lower outflow velocity. In particular, we do not know whether the kinetic energy of the wind
is sufficient for the gas to escape from the galaxy. A simple estimate of the escape velocity obtained in terms of
the isothermal sphere model with a cutoff radius of  2 arcmin yields $v_{esc}\geq2.7v_{max}=180\km$ for regions located at
galactocentric distances $r<8$ arcsec, which is appreciably greater than $v_{out}$. Therefore, most probably, after cooling
down, the gas of the wind will fall back onto the galactic plane.

\subsection{Cause of the burst of star formation}

Thus  the central kiloparsec in NGC4460 is engulfed in on-going  star formation
\citep[$SFR=0.3\,M_\odot/yr$, see][]{KaisinKarachentsev2008}, in contrast to the rather old and  `quiet' disc of the
galaxy.  What could have triggered star formation in this lenticular galaxy? Tidal interaction with a companion is unlikely since the galaxy appears  isolated. Neither there is any evidence for a merger of a gas-rich dwarf companion -- the velocity field lacks kinematically decoupled components to be expected in such a case. No tidal features can be seen on optical images or published low-resolution HI maps~\citep{Sage2006}. The inner disc that we found in the brightness distribution proves that star formation has been going on for quite a long time in the center, however it  is not indicative of interaction.

We already pointed out above that isolated E and S0 galaxies are rather rare objects,  their integrated
parameters differ appreciably from normal E and S0 galaxies in groups and clusters. Table~\ref{tab2} lists
the basic parameters of all the three representatives of this category of galaxies found within 10~Mpc from the Sun.
Row (2) gives the tidal index according to \citet{Karachentsev2004}; distances in row (3) are from \citet{Tonry2001} for except NGC~404 for which mean value from \citet{Tonry2001} and \citet{Karachentsev2002} is accepted;
Rows (1), (4) and (5) give the observed parameters of the galaxies adopted from the  NED database; Row (6) presents the HI mass from HyperLEDA database adopted for the accepted distances. Row (7) gives the estimated global star-formation rate in galaxies inferred from their H$\alpha$ fluxes  from  \citet{KaisinKarachentsev2008} for NGC~4460 and from \citet{KarachentsevKaisin2010} for other two galaxies.  The last rows give two dimensionless and distance independent parameters (see \citet{KarachentsevKaisin2007} and \citet{KaisinKarachentsev2008} for details): $P^*_K = \log([SFR]\times T_0/L_K)$ and $F^*=\log (M_{HI}/[SFR]\times T_0)$. The former characterizes the proportion
of its luminosity the galaxy would produce during the Hubble time $T_0=13.7$ Gyr  at the current rate of star formation and the mass-to-(K-band)-luminosity ratio $M/L_K = 1M_{\odot}/L_{\odot}$. The  parameter $F^*$ shows how much Hubble time the galaxy will need in future to spend the present supply of gas if star formation proceeds at the currently observed rate.

\begin{figure}
\centering
\includegraphics[width=8.8cm]{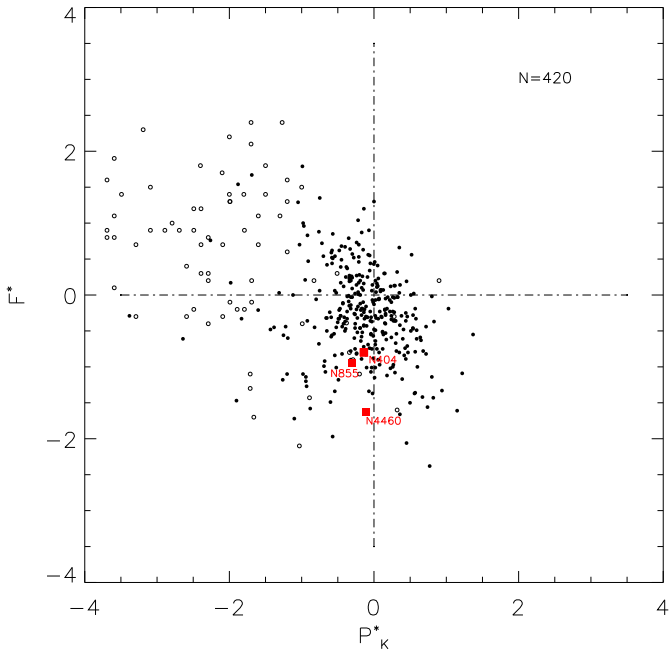}
\caption{Evolutional plane `past-future' for 420 Local Volume galaxies \citep{KarachentsevKaisin2010}. The galaxies observed and detected in H$\alpha$ are shown by filled symbols, and galaxies with only upper limit of their H$\alpha$ flux are indicated by open circles. The isolated E/S0 galaxies of moderate luminosity are distinguished by red squares and marked  with their NGC numbers.}
\label{fig7}
\end{figure}

Distribution of 420 Local volume galaxies in the `past-future' diagram  \citep{KarachentsevKaisin2010} is presented in Fig.~\ref{fig7}.
The discussed isolated early-type galaxies of moderate luminosity are distinguished on the diagram by red color. Here, galaxies of different morphological types occupy essentially different regions in the diagnostic diagram. The median values of $P^*_K$ and $F^*$ for the total LV sample are $-0.40$ and $-0.25$, respectively. These quantities mean that a typical LV galaxy requires $2-3$ times higher SFR in the past to reproduce its
observed luminosity, having also enough amount of gas to continue star formation with the observed rate during the next 8 Myrs.

The NGC~404, NGC~855, and NGC~4460 galaxies differ appreciably in luminosity, but have almost similar
hydrogen masses $\sim 1\times 10^8M_{\odot}$. Specific star-formation rate per unit luminosity differs little
among these galaxies. The parameter $P^*_K$ is only slightly smaller than zero for all three galaxies, and this
indicates that the galaxies are capable of reproducing most of their stellar mass during cosmological time $T_0$
provided that star formation is at the current level. At the same time, the available gas reserves in these galaxies can maintain the current rate of star formation only for a short time scale ranging from 1/40 (NGC~4460) to 1/6 (NGC~404) of the age of the Universe.

The location of isolated E and S0 galaxies on the diagnostic diagram $\{P^*_K,F^*\}$ does not allow us to view
their observed evolutionary stage as a burst of star formation triggered by a merger of a companion or a
single hydrogen cloud. Should this be the case, one would expect in the Local Volume, in addition to the three
isolated emission E and S0 galaxies mentioned above, an order of magnitude greater number of similar objects at
the quiescent interburst stage. However, the currently available \Ha-flux data for 420 galaxies of the Local Volume \citet{KarachentsevKaisin2010} do not support the existence of such a population. Besides NGC~4460, NGC~404, and NGC~855, the Local Volume contains only two other isolated galaxies: NGC~2787 (S0-a) and NGC~4600 (S0) where very weak  \Ha\, emission is presented. These galaxies gave blue absolute magnitudes $-18.5$  and $-15.7$, respectively.

The predominance of objects with current central star formation among isolated E and S0 galaxies and their capability of reproducing the available stellar mass over the cosmological time scale   with the observed SFR  combined with the scarcity of available gas reserves ($\sim 1\times 10^8M_{\odot}$) leads us to suggest that ongoing star formation in this galaxies  is fed by an external source, i.e. intergalactic gas clouds or filaments. The process of external gas accretion  should be have  a regular steady character on a cosmological time scale, without significant bursts.

Recently \citet{Kannappan2009} described  a population of `blue-sequence E/S0s' resided on the locus of late-type  galaxies  in color vs. stellar mass parameter space. They argue that these galaxies actively (re)growing their stellar discs and therefore may form a bridge between  late-type spirals/iregulars and early-type red-sequence galaxies. Many of low-to-intermediate mass blue-sequence E/S0s reside in low density environments, often have a blue center and substantial fraction of cold gas. However their gas consumption time usually smaller than 3 Gyr \citep{Wei2009}. All these properties are similar with  the Local Volume isolated E/S0 galaxies described above, therefore  NGC~4460, NGC~404, and NGC~855 seem to be belonging to the population of blue-sequence E/S0s. \citet{Kannappan2009} suggest these galactic population connects with the  disc rebuilding  after major/minor merging or companions interactions. In the present paper we stress that accretion of external gas is also important in a  particular case of the evolution of isolated early-type galaxies.


\begin{table}
\caption[]{Integrated parameters of isolated E/S0 galaxies within 10 Mpc}
\label{tab2}
\centering
\begin{tabular}{rlrrrrr}
\hline\hline
Line & Parameter             & NGC~4460   & NGC~855 &  NGC~404   \\
\hline
(1)  & Type                 &   SB(s)0+     &     E         &  SA(s)0-   \\
(2)  & TI                   &  -0.7         &   -0.8        &    -1.0    \\
(3)  &Distance  (Mpc)       & 9.59          & 9.73          &   3.24     \\
(4)  &M$_B$      (mag)      & -17.89        &  -17.07       &   -16.61   \\
(5)  &M$_K$      (mag)      & -20.87        &  -20.15       &   -19.00   \\
(6)  &$\log$M(HI)  (sun)    & 7.92          &    8.13       &     7.97   \\
(7)&$\log$[SFR](M$_\odot$/yr)&  -0.59         &   -1.07       &   -1.37    \\
(8)  &$P^*_K$               &  -0.11        &   -0.30       &    -0.14   \\
(9)  &$F^*$                  &  -1.63        &   -0.94       &    -0.80   \\
\hline
\end{tabular}
\end{table}

\section{Summary}

We used available UV and optical images  to explore the
structure of the isolated lenticular galaxy NGC~4460. Also we obtained long-slit and 3D spectroscopic data to probe the origin of the bright \Ha\,nebulosity surrounding the central region of the galaxy.   We consider the following points to be the most important.

\begin{enumerate}
\item All ongoing star formation with the rate about $0.3\,M_\odot/yr$ resides in a pseudo-ring with radius about one kiloparsec. Together with antitruncated shape of the surface brightness profile it suggests a recent formation of the inner exponential disc, whose radial scale three times smaller than global old stellar disc.
\item Whereas gas in the circumnuclear disc is photoionized  by radiation of young stars, the external regions of \Ha\,nebulosity are ionized by shocks. This fact, in combination with a bi-conical shape of the ionized gas nebulosity, with  the radial distributions of the emission lines velocities and width can be explained  as a galactic-scale outflow (superwind) induced by a central star formation. We estimated outflow velocity as $\geq130\km$. Galactic wind parameters, such as outflow velocity,  formation time and  total kinetic energy are several times smaller comparing with the known galactic wind  in  NGC\,253, which is explained by substantially different star formation rates.
\item Also we discussed the evolutional status of NGC~4460 together with other well-isolated E/S0 galaxies in the local volume: NGC~404, and NGC~855. We considered the position of the galaxies on the `past-future' diagram \citep{KarachentsevKaisin2010} and suggested that continuous accretions of  an intergalactic gas is most probable source of their current star formation.  In contrast to merging scenario, the external gas accretion seems to be  a sluggish process without violent events.

\end{enumerate}

The research is partly based on data obtained from the Multimission Archive at the Space Telescope Science Institute (MAST). STScI is operated by the Association of Universities for Research in Astronomy, Inc., under NASA contract NAS5-26555. Support for MAST for non-\textit{HST} data is provided by the NASA Office of Space Science via grant NAG5-7584 and by other grants and contracts. Funding for the SDSS and SDSS-II has been provided by the Alfred P. Sloan Foundation, the Participating Institutions, the National Science Foundation, the US Department of Energy, the National Aeronautics and Space Administration, the Japanese Monbukagakusho, the Max Planck Society and the Higher Education Funding Council for England. The SDSS web site is http://www.sdss.org/. This research has made use of the NASA/IPAC Extragalactic Database (NED) which is operated by the Jet Propulsion Laboratory, California Institute of Technology, under contract with the National Aeronautics and Space Administration. We acknowledge the usage of the HyperLeda database (http://leda.univ-lyon1.fr).
This work was supported by the Russian Foundation for Basic Research (projects no.~09-02-00870, no.~07-02-00005, and no.~08-02-00627). AM is also grateful to  `Dynasty' Fund. The authors thank the anonymous referee for his constructive advice that helped us to improve the paper.

\label{lastpage}
\end{document}